\documentclass[10pt]{article}

\usepackage{amsmath}
\usepackage{amssymb}

\usepackage{graphicx}

\usepackage{cite}

\usepackage{color}

\usepackage{booktabs}

\usepackage{setspace} 
\usepackage[labelfont=bf,labelsep=period,justification=raggedright]{caption}

\makeatletter
\renewcommand{\@biblabel}[1]{\quad#1.}
\makeatother

\date{}

\usepackage{verbatim}
  \usepackage{lineno}
  \linenumbers

\begin{document}
\begin{flushleft}

{\Large
\textbf{Detecting range expansions from genetic data}
}
\\
Benjamin M Peter$^{1,\ast}$, 
Montgomery Slatkin$^{1}$, 
\\
\bf{1} Department of Integrative Biology, University of California, Berkeley, Berkeley, California 94720-3140, USA
\\
$\ast$ E-mail: bp@berkeley.edu
\end{flushleft}

\section*{Abstract}
We propose a method that uses genetic data to test for the occurrence of a recent range expansion and to infer the location of the origin of the expansion. We introduce a statistic for pairs of populations $\psi$ (the directionality index) that detects asymmetries in the two-dimensional allele frequency spectrum caused by the series of founder events that happen during an expansion. Such asymmetry arises because low frequency alleles tend to be lost during founder events, thus creating clines in the frequencies of surviving low-frequency alleles. Using simulations, we further show that $\psi$ is more powerful for detecting range expansions than both $F_{ST}$ and clines in heterozygosity. We illustrate the utility of $\psi$ by applying it to a data set from modern humans and show how we can include more complicated scenarios such as multiple expansion origins or barriers to migration in the model.

\section*{Author Summary}
Many important biogeographic processes can be interpreted as range expansions, where a species is constraint to a small local habitat in the beginning and expands from there to colonize a larger region. Examples of this include biological invasions, the spread of infectious diseases and humans colonizing the world. We present a statistical framework to test for the existence of such an expansion and examine its properties. We then use this framework to make further inference, and show how it can be used to estimate the origins of the range expansion and to identify barriers of gene flow.
\section*{Introduction}
Range expansions are ubiquitous in natural populations, and they are responsible for numerous biological phenomena. Range expansions result in a series of founder events that cause the newly founded populations to differ genetically from the source population. Some well-known examples are biological invasions\cite{handley2011}, the post-ice age patterns of migration in several European \cite{hewitt1999,schmitt2007}, and the colonization of Eurasia and North and South America by modern humans\cite{cavalli-sforza1994,ramachandran2005,tishkoff2009}. In some cases descendants from the source population remain near the location of the ancestral population. For example, the European population of the brown bear \textit{Ursus arctos} most likely survived the last ice age in refugia in Spain and Greece. Brown bears followed the receeding glaciers to colonize most of Europe, but populations at the locations of the former refugia persisted until the populations were driven to the verge of extinction by humans in the 
20th century\cite{taberlet1998}. Another example are humans, where derived populations are found all over the world, but there are also descendants of the first humans still living in Africa.

Sometimes, the routes of migration are known from direct observations, historical records and archaeological evidence. Frequently, however, the exact history of a species is unknown, and we want to use population genetic methods to gain more information. In this paper, we use genetic data to address two related problems: detecting whether a range expansion has occurred and inferring the geographic origin of a range expansion.

Characterizing the influence of geographic structure on genetic diversity has been one of the major goals of population genetics theory, with important contributions from Wright\cite{wright1943}, Mal\'{e}cot\cite{malecot1950}, Kimura\cite{kimura1964} and many others. While there are many statistics designed to infer differentiation between populations\cite{goldstein1995, nei1972,reynolds1983,balakrishnan1968}, the most widely used statistic to detect differentiation between populations is the fixation index $F_{ST}$, which traces to Wright \cite{wright1949}. A variety of estimators of $F_{ST}$ have been developed (e.g \cite{reynolds1983,weir1984}). Roughly speaking, $F_{ST}$ measures how much diversity exists between subpopulations compared to the diversity in the entire population; a value of 0 indicates that the two subpopulations are indistinguishable,  whereas a value of 1 indicates that two populations are maximally differentiated. $F_{ST}$ has been directly linked to the 
migration rate in several models, including the finite island\cite{slatkin1991} and stepping-stone models\cite{cox2002}. Although $F_{ST}$ can be used to quite accurately estimate the amount of gene flow between equilibrium populations, it cannot be used to infer directionality of gene flow. 

Two other methods that are widely used to detect geographic patterns are clustering algorithms and ordination methods. Clustering analyses \cite{pritchard2000,corander2004,francois2008,francois2010} such as STRUCTURE \cite{pritchard2000}) classify individuals into discrete groups, which can then be used for further analysis. Ordination techniques\cite{cavalli-sforza1967}, such as principal components analysis and multidimensional scaling summarize data by indicating the overall similarity of populations. Principal component analysis has been shown that genetic diversity relatively closely correlates to the geographic distribution of humans on a continental \cite{novembre2008} and global \cite{cavalli-sforza1996,wang2012} scale. 

It is also possible to use likelihood methods to infer past features of population history. For example, the program IM \cite{Hey2010} estimates the time of separation of populations and migration rates between them using data from multiple unlinked loci, and the program dadi \cite{gutenkunst2009} estimates past rates of population growth from the joint allele frequency spectrum from two or three populations. Both of these programs are computationally intensive and neither can analyze data from numerous populations.

Most statistics applied to subdivided populations do not provide information about asymmetries. $F_{ST}$ and most genetic distances are defined in such a way that they are commutative (i.e. $F_{ST}$ between populations A and B is the same as $F_{ST}$ between B and A), and hence the value depends only on the amount of migration, not whether migrants moved mostly from  A to B or from B to A. Clustering algorithms can produce groupings of populations that can be interpreted as describing an expansion, but the expansion-specific information is lost in the process and the results of clustering is often sensitive to tuning parameters such as the number of clusters. For principal components analysis, the view that the first principal component axis follows the direction of expansion \cite{menozzi1978} has recently been challenged \cite{novembre2008,francois2010,degiorgio2012}, and it has recently been shown that, depending on parameters and the locations of the populations sampled, the first principal component 
axis might be parallel to or orthogonal to the axis of expansion, or at an angle in between.

Population genetics theory has shown that a range expansion can be detected from the characteristic reduction in genetic diversity with increasing distance from the origin of the expansion\cite{austerlitz1997,edmonds2004,ramachandran2005,hallatschek2007,degiorgio2009,slatkin2012}. The reason is that the succession of founder events during the expansion cause the progressive loss of genetic variants. This prediction has been confirmed by comparing the numbers of mtDNA haplotypes  found in Southern European refugia and in central Europe \cite{taberlet1998}. The same pattern can also been seen in humans where both a reduction in heterozygosity  and an increase in linkage disequilibrium with increasing distance from the presumed origin of the expansion in Africa can be shown \cite{ramachandran2005}.

In addition to creating a gradient in genetic diversity, range expansions tend to create clines in the frequencies of neutral alleles, with the frequency increasing on average in the direction of the expansion \cite{slatkin2012}. An intuitive reason for this pattern is that each founder event results in additional genetic drift, and populations further away from the origin of expansion will therefore have experienced more genetic drift. This can be seen from the following simple argument: The expected frequency of an neutral allele in the new population is the same as in the source population. But some alleles will have zero frequency in the new population. Therefore, the average frequency of shared alleles, i.e. alleles that are present in the new population, is expected to be higher than in the source population, thus creating the cline. This observation provides the foundation for our method of detecting range expansions.

In this paper, we introduce a statistic, the directionality index $\psi$, for pairs of populations. $\psi$ is sensitive to patterns created by range expansions as it detects clinal patterns created by successive range expansions. We show, using simulations, that the expectation of $\psi$ is zero in an equilibrium isolation-by-distance model, and that positive values of $\psi$ indicate the direction of the expansion. We also show that if we have multiple samples, $\psi$ can be used to infer the origin of a range expansion and the location of barriers of gene flow. We also explore the power and robustness of our methods and finally apply it to human genetic data.

\section*{Results}

In this section, we will define the directionality index, give an intuitive explanation and discuss some of its properties. We will show that the directionality index is sensitive to recent range expansions in a one or two dimensional stepping-stone model, and then explore some more advanced applications.
\subsection*{Definition Of The Directionality Index}
Consider two samples of size $n, n \geq 2$ taken from two subpopulations $S_1,S_2$. Each sample consists of $L$ biallelic markers (e.g. SNPs) that are shared between $S_1$ and $S_2$. The directionality index is defined as 
\begin{equation}
 \psi = \frac{1}{n} \left( \bar{f}_{S_1}-\bar{f}_{S_2} \right) = \frac{1}{L n} \sum_{l=1}^{L}\left( f_{S_1,l}-f_{S_2,l} \right), \label{dIndex}
\end{equation}
where $\bar{f_S}$ is the average allele frequency of all alleles in population $S$, and $f_{S,l}$ is the number of derived copies of allele $l$ in the sample from population $S$.  Equivalently, $\psi$ can also be defined in terms of the  two-dimensional site frequency spectrum (2D-SFS): 
\begin{equation}
 \psi = \sum_{i=1}^{n}\sum_{j=1}^{n}(i-j)f_{ij} \label{dIndexAlt}.
\end{equation}
where $f_{ij}$ denotes the proportion of SNP in the sample that are at frequncy $i$ in $S_1$ and at frequency $j$ in $S_2$, and the SFS is normalized such that $\sum_{i<j}f_{ij}=1$.
\begin{equation}
 \psi = f_{21} - f_{12}\text{.} \label{dIndexSize2}
\end{equation}
 The three different definitions represent different interpretations of the directionality index, and it is useful in building an intuition to discuss them briefly: Equation \ref{dIndex} corresponds to the definition alluded to in the introduction, where we compare the average allele frequency between the two populations. As the population further away from the expansion origin is expected to have experienced more genetic drift, its alleles are expected to be at a higher frequency and thus $\psi$ is positive if $\bar{f}_{S_1} > \bar{f}_{S_2}$ and $S_1$ is further away from the origin of the expansion. If both populations have experienced similar amounts of genetic drift, then the allel frequencies will be equal, $\psi\approx 0$ and we will not detect an expansion. Equation \ref{dIndexAlt} is based on the SFS, and we see that $\psi$ will be positive if $f_{ij}$ is usually greater than $f_{ji}$. Thus, we are comparing the SFS entries that are mirrored along the $x=y$ diagonal, and the directionality index 
measures the ``skew'' in the 2D-SFS. If there are more SNP that fall in the upper left triangle of the SFS (where $j>i$), $\psi$ will be negative, and we infer an expansion from $S_1$ to $S_2$. The inverse conclusion will be drawn if there is an excess of SNP in the lower right triangle, and if the SNP are distributed symmetrically around the $x=y$ diagonal, $\psi$ will be zero. Much of the paper will be focused on the case where each population is represented by a single genome, a case we think will be particularly common in many analyses. In this case, equation \ref{dIndexAlt} reduces to \ref{dIndexSize2} and we are simply comparing the abundance of SNP that are fixed for the derived allele in sample $S_1$ and heterozygous in $S_2$ to the number of SNP that are heterozygous in $S_1$ and fixed in $S_2$. If either number is significantly larger than the other, we infer migration in the direction of the larger number. 
 It is also worth noting that the computation cost of equation \ref{dIndex} scales proportionally to the number of loci in the sample, whereas equation \ref{dIndexAlt} only depends on the sample size squared. Thus, for genome scale data sets where $L>>n^2$, (\ref{dIndexAlt}) will be much faster to compute.

\subsubsection*{Determining Whether A Range Expansion Occurred}
\begin{figure}[!ht]
\begin{center}
\includegraphics[width=6in]{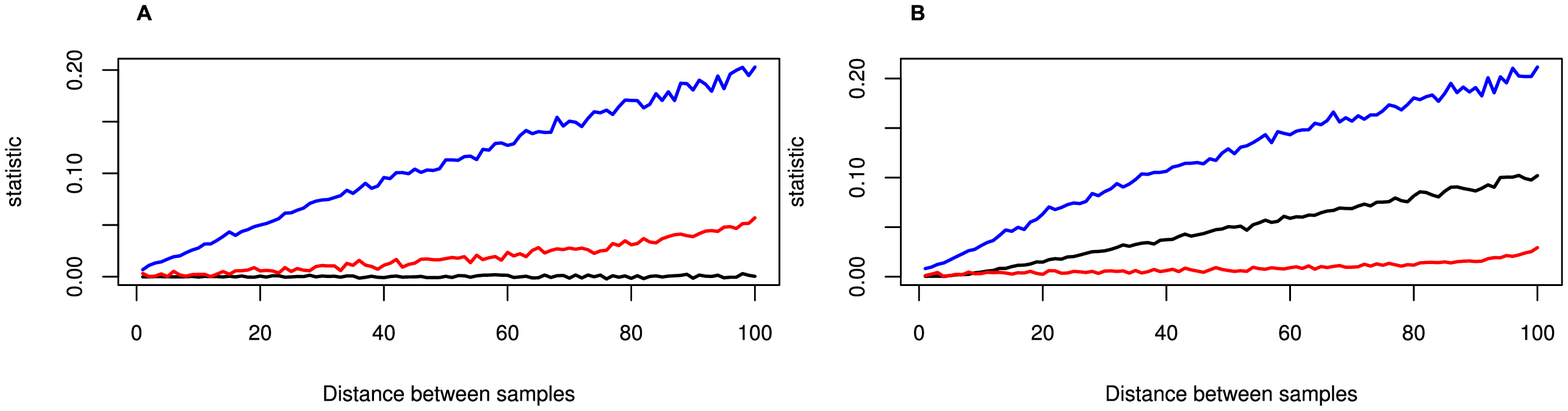}
\end{center}
\caption{
{\bf Behavior of $H$ (red), $\psi$ (black) and $F_{ST}$ (blue) in one-dimensional (A) isolation-by-distance and (B) population-expansion models.} Simulations were performed on a 200-stepping stone model with scaled migration rate $M$=100 between adjacient demes, and expansion events every 0.001 coalescence units. $F_{ST}$ increases linearly with distance in both models and $\psi$ is zero in the isolation-by-distance model, but increases approximately linearly in the expansion model. Heterozygosity increases from the center of the population (left) to the border of the habitat (right).
}
\label{FigDistance}
\end{figure}

We first test the power of $\psi$ to distinguish pairs of populations sampled from a recent range expansion to pairs of populations sampled under isolation-by-distance at equilibrium in a 1D-model. Figure \ref{FigDistance} shows that $F_{ST}$ increases at approximately the same rate under an equilibrium stepping-stone model with only isolation-by-distance (Panel A) and a model with a range expansion (Panel B), indicating that the two scenarios are comparable. We see that $\psi$ is constant at zero in the isolation-by-distance model, regardless of the distance between the samples. In contrast, $\psi$ increases with distance under the expansion model, due to the increase in allele frequency along the expansion axis. Interestingly, $\psi$ increases almost linearly with the distances between the origin and the population sampled, a fact we exploit in the next section to infer the origin of the expansion. We also plotted the heterozygosity, a statistic that is also expected to be constant under an equilibrium 
model \cite{durrett2008} and increasing under an expansion \cite{austerlitz1997,ramachandran2005}. However, our simulations show that heterozygosity is larger in the center of the habitat than near the boundaries because of the boundary effects. This is in contrast to most theoretical results \cite{durrett2008} which either assume either a circular model or an infinitely long stepping stone model, and where the heterozygosity is independent of the sampled deme. However, the observed gradient in heterozygosity has been observed previously and explained by longer coalescence times for a sample taken close to the boundary \cite{maruyama1970,wilkins2002} and it is worth noting that this effect is much weaker in a two dimensional population.

\begin{figure}[!ht]
\begin{center}
\includegraphics[width=6in]{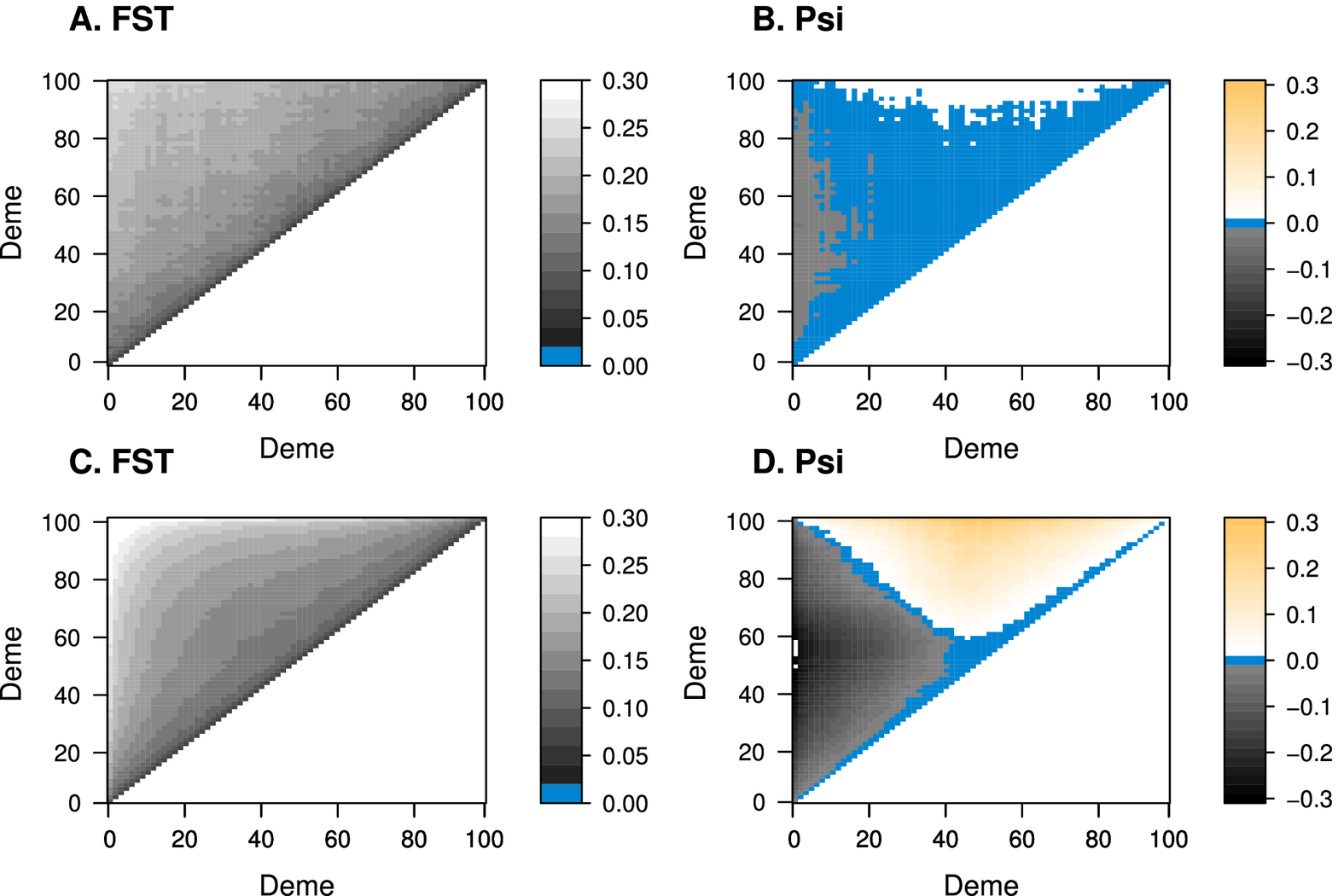}
\end{center}
\caption{
{\bf Behavior of $F_{ST}$ and $\psi$ in isolation-by-distance and population expansion model.}  Each panel gives the value of the pairwise statistics $F_{ST}$ (Panels A and C) and $\psi$ (Panels B and D) under an isolation-by-distance model (Panels A and B) and an expansion model (Panels C and D) starting in the central deme (50,50). Simulations were performed on a 101 x 101 deme stepping stone model, and a diagonal transect from demes at coordinates (0,0) to (100,100) was sampled, and all pairwise statistics were calculated. Blue regions correspond to regions where $F_{ST}$ and $\psi$ are very low (below 1\%). The orange and grey regions denote areas with positive and negative $\psi$, respectively. An  Whereas $F_{ST}$ behaves qualitatively similar under both models, the behavior of $\psi$ is very different. Under isolation-by-distance, $\psi$ is very close to zero, with some deviations due to boundary effects. Under an expansion, however, we see a clear signal of an expansion for all demes, except demes 
that are very 
close to each other, or demes that have the same distance to the origin, but in different directions.
}
\label{FigLevelplot}
\end{figure}

On the other hand, $F_{ST}$ and $\psi$ behave in similar ways in both 1D and 2D models (Figure \ref{FigLevelplot}). $F_{ST}$ is slightly larger in the case of a range expansion than in the isolation-by-distance model (Panels A and C), but qualitatively we see an increase of $F_{ST}$ with distance under either model. The pattern for $\psi$, however, is again different (Panels B and D): under the isolation-by-distance model, $\psi$ is smaller than 0.01 for almost all comparison, with the exception of a few demes that are at the boundary of the simulated region. In contrast, the magnitude and sign of $\psi$ nicely illustrate the effect of the range expansion. $\psi$ is zero only for demes that are very close to each other or demes that are equally far away from the expansion origin. The latter can be explained by symmetry: two samples that are an equal distance apart from the origin will have a symmetric SFS, resulting in a $\psi$ close to zero. 

Various significance tests can be used to determine the significance of $\psi$ between two populations; for the case of $n=2$ in both samples we can simply perform a binomial test on the absolute frequencies $f_{21}$ and $f_{12}$. If their proportions differ significantly from 0.5, we can reject the null hypothesis of symmetric migration between the two demes. When comparing samples of size $n>2$, we can generate a null distribution using a permutation test, i.e randomly assigning the allele frequencies for each SNP to either population. However, both these tests will underestimate the variance in the data if SNPs are not in linkage equilibrium. In that case the ``effective'' number of loci will be lower than the actual number. To take linkage into account we use a computationally more intensive block-jackknife approach \cite{busing1999,reich2009} to analyse our real data.

\begin{figure}[!ht]
\begin{center}
\includegraphics[width=6in]{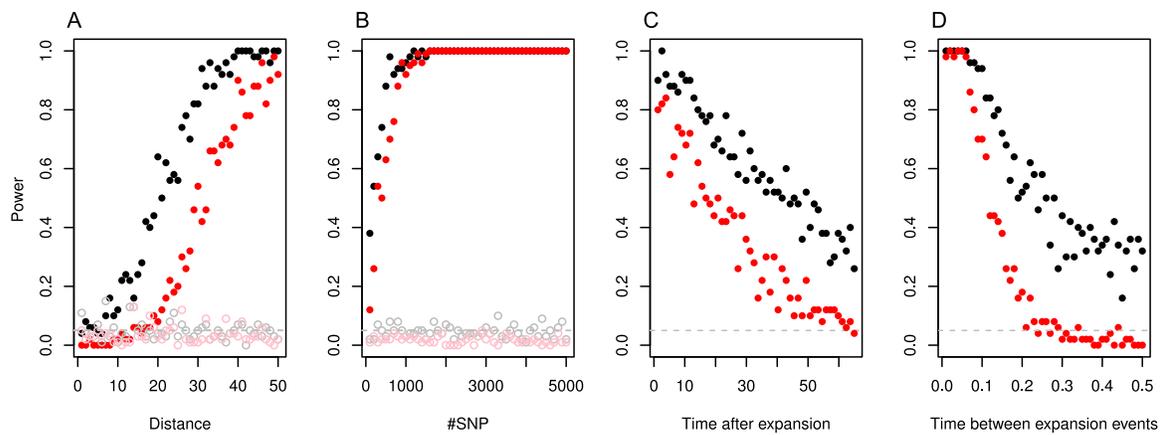}
\end{center}
\caption{
{\bf True/false positive rates of detecting range expansion}  Each panel give the proportion of replicates in which the null model was rejected at the 5\% significance level. Filled dots denote simulations under an expansion model, and open dots correspond to an isolation-by-distance model. Black dots corresponds to using the directionality index, and the red dots were generated using the difference in heterozygosity as a statistic. The grey dashed line at 0.05 gives the expected proportion of false positives under the null hypothesis. Baseline parameters for the simulations were of 2 chromosomes (one diploid individual) at each location sampled, with locations a distance of 50 each other. Each sample consisted of 1,000 independent SNPs shared between the two populations. Significance was assessed using a binomial test.
}
\label{FigPowerIBD}
\end{figure}

In Figure \ref{FigPowerIBD} we show the effect of the most important parameters on our ability to reject the null hypothesis of isolation-by-distance for pairs of samples of size two. For all parameters, we find that using the directionality index results in higher power than comparing differences in heterozygosity, while false-positive rates are low and roughly the same for the two methods. We find that we have comparatively little power to reject the null hypothesis if the two sampled individuals are close to each other(Panel \ref{FigPowerIBD}A). This is expected, since there are fewer founder events separating the two individuals. Therefore we expect $\psi$ to be lower for nearby populations, as shown in Figures \ref{FigDistance} and \ref{FigLevelplot}. Panel B shows that a moderate number of shared SNPs is necessary, i.e. more than one thousand, to get high power to reject equilibrium isolation-by-distance. In addition, we find that slow expansions are harder to detect than rapid expansions, and more 
recent expansions are easier to detect than expansions that happened 
a long time in the past (Panels C and D). Neither of these findings is unexpected; after an expansion genetic drift will affect the loci in both populations equally. The number of shared SNP that are due to the range expansion will decrease with time and be partially replace by SNP that only experienced the equilibrium model and hence do not carry a signal of the expansion. Similarly, if the time between expansion events is high, the founder effects caused by the expansion will become less important relative to genetic drift between expansion events, weakening the signal of the expansion. In this scenario, the power of heterozygosity to detect an expansion decays much faster than the power of $\psi$. Finally, we note that the false positive rate, denoted in grey and pink in Figure \ref{FigPowerIBD}, is independent of both the distance between loci and the number of SNPs.

\subsection*{Inferring The Origin Of A Range Expansion}
 In addition to showing that range expansion occurred, the results in Figures 1 and 2 suggest that spatial patterns in pairwise values of $\psi$ can indicate the origin of an expansion if we have more than two samples. For this purpose, we employ a method commonly used by engineers in problems of localization and navigation\cite{Gustafsson2003}, called Time Difference of Arrival location estimation (TDOA). TDOA methods are used in remote sensing and to locate cell phones \cite{Gustafsson2003}. The key assumption of the TDOA algorithm is that the magnitude of a pairwise statistic between two sample locations $i$ and $j$ is proportional to the difference in distance from $i$ to the origin and the distance from $j$ to the origin. So, if $i$ is very close to the origin and $j$ far away, then we would expect the TDOA statistic to be large, but if $i$ and $j$ are at the roughly the same distance from the origin, then the TDOA statistic should be close to zero. In engineering applications the TDOA statistic is the 
time difference between the arrival of a signal 
emitted from the origin (hence the name). In our application, however, $\psi$ takes on the role of the time difference with the implicit assumption that the directionality index between locations $i$ and $j$, $\psi_{ij}$, is proportional to the difference in distances from $i$ and $j$ to the origin, respectively. To further illustrate how we can use this method to infer an origin, we first consider the special case of $\psi_{ij}=0$. Assuming that we have already rejected isolation-by-distance, we know that $i$ and $j$ are equally far from the origin and the origin must therefore lie on the line perpendicular to the line through $i$ and $j$. 

\begin{figure}[!ht]
\begin{center}
\includegraphics[width=6in]{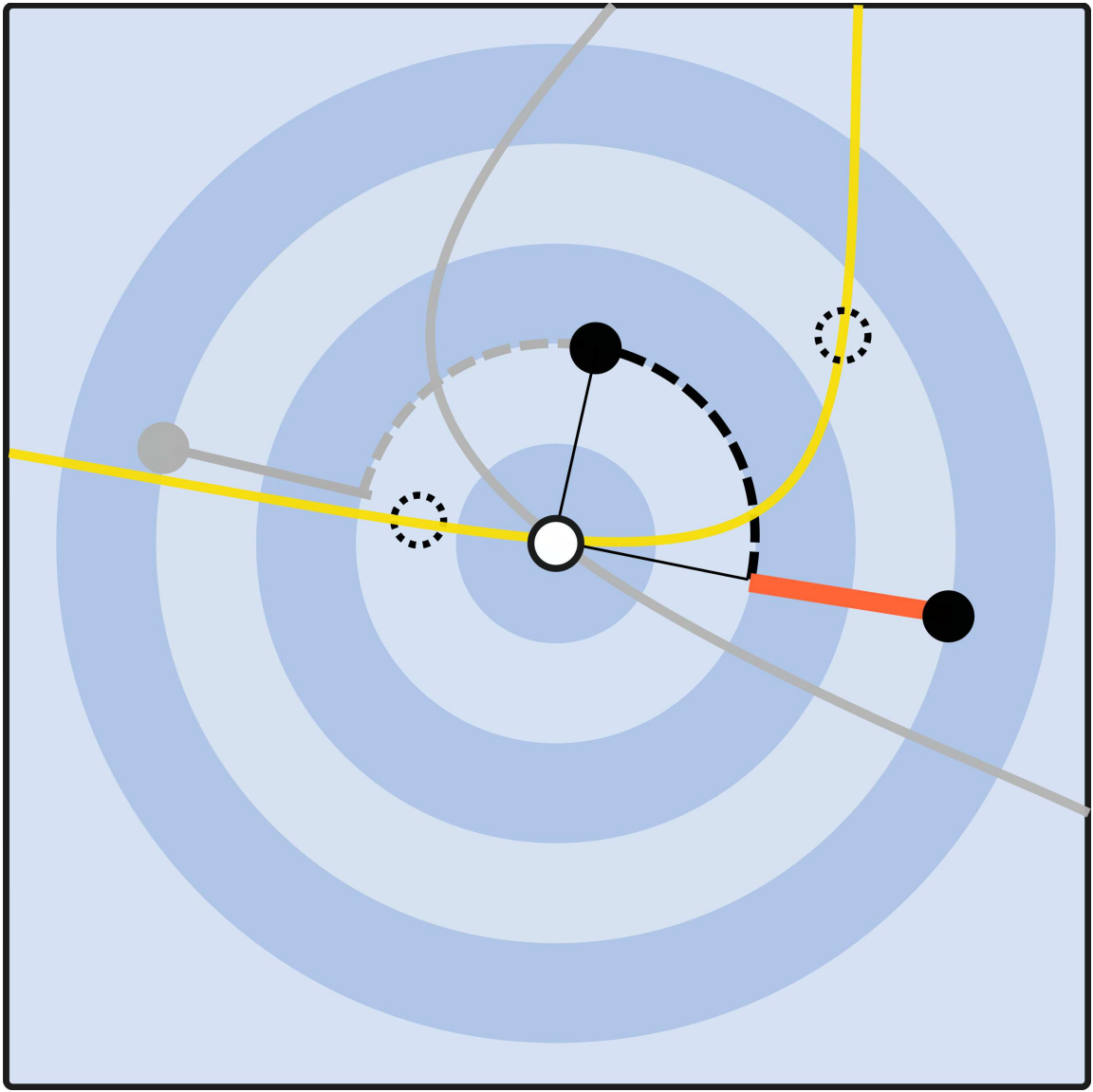}
\end{center}
\caption{
{\bf Illustration of the method used to infer the origin of a range expansion.} The black and grey points correspond to genetic samples taken, the white point corresponds to the (unknown) origin of the expansion. Using the directionality index $\psi$, we can infer the difference in distance from the samples to the origin (orange line). The set of all points that has the same difference in distance to the origin corresponds to the arm of a hyperbola (yellow), with some candidate points outlined with the dashed point. Using a second pair of points (the grey and top black point), we can identify a second hyperbola, and find an unique location of the origin. In practice, we use more than three sampling locations. Sampling noise will cause the hyperbolas to not intersect in a single point. We use a least-squares criterion to estimate the location of the origin.
}
\label{fig.tdoaCartoon}
\end{figure}

If we had three or more loci all at the same distance from the origin (so that the pairwise $\psi$ values are all zero), we could thus infer the origin as the center of the circle passing through the three points. In general, however, $\psi$ will be non-zero. In that case, we know from elementary geometry that the set of candidate points based on a one pair of samples is not a straight line, but a hyperbola with $i$ and $j$ as its focal points. If we have samples from $k$ locations, we can calculate $\psi$ for $k (k-1) /2 $ samples and hence obtain $k*(k-1)/2$ hyperbolas. In a perfect, noiseless world, all hyperbolas would intersect in a single point: the origin of the expansion, as illustrated in Figure \ref{fig.tdoaCartoon}. In practice, of course, genetic data is highly stochastic and we have to estimate the origin. To do this, we interpret each hyperbola as a non-linear equation with three unknowns, the sample coordinates $x,y$ and the speed of expansion $v$. $v$ is a nuisance parameter that describes 
how much the allele frequency increases per unit distance from the origin. For more than three samples the system is overdetermined and, rather than solving the system of equations explicitly, we  use weighted non-linear least squares.

\begin{figure}[!ht]
\begin{center}
\includegraphics[width=6in]{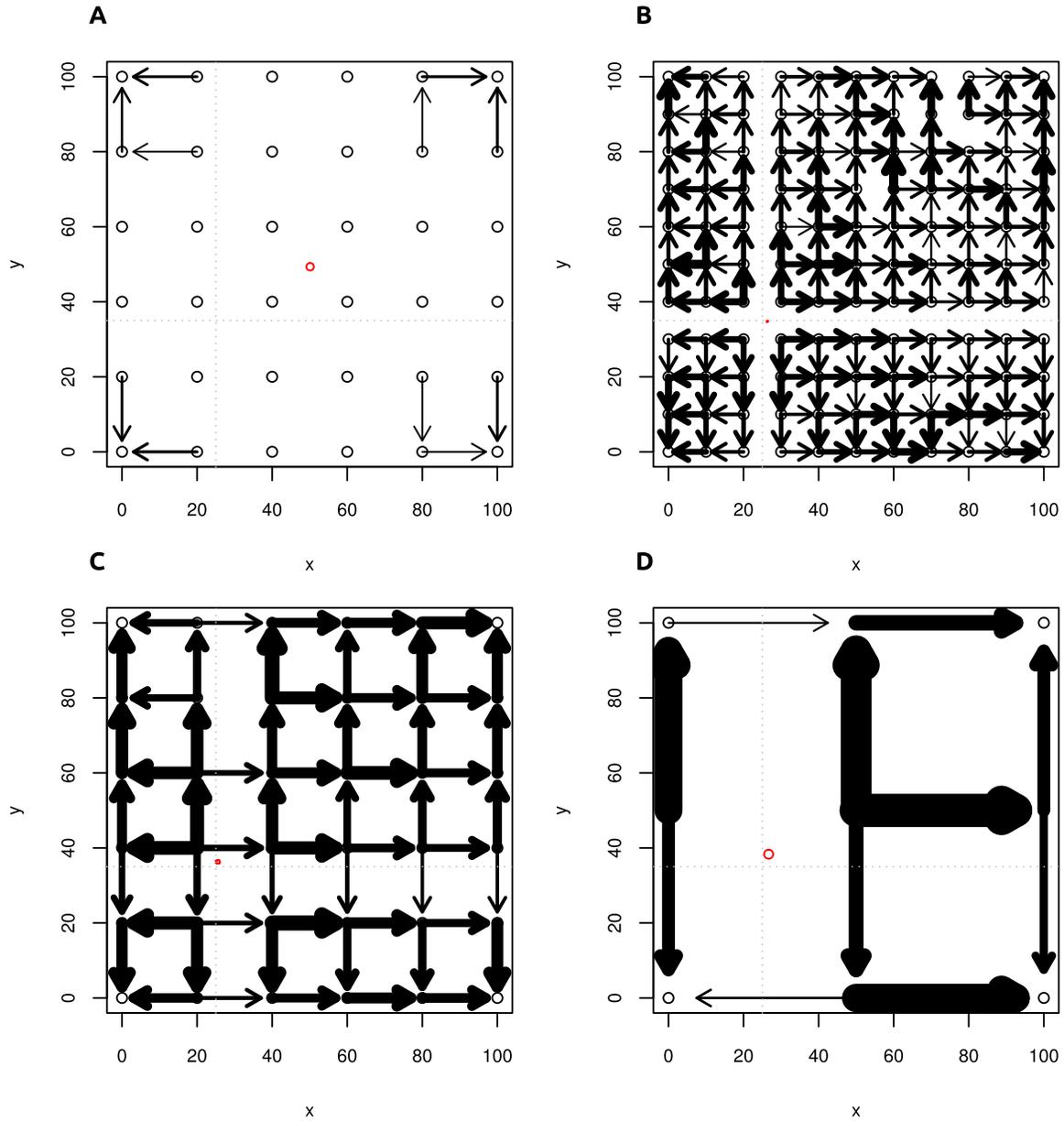}
\end{center}
\caption{
{\bf Detecting the origin of a range expansion.} Each panel corresponds to a 101 x 101 grid of populations that were simulated. The expansion began at point (25,35) (indicated by gray dotted lines). Black bordered circles indicate sampling locations, black arrows correspond to $\psi>1\%$ between adjacent samples, with the direction of the arrow indicating the sign of $\psi$. Thicker arrows correspond to larger $\psi$. The red ellipse corresponds to the 95\% confidence interval of the estimated location of the origin. Panel a: no expansion (isolation-by-distance model). Edge effects cause the estimated origin to be close to the center of the grid of populations. Panels b-d: Expansion with parameters $M=1$, $t=0.1$ and samples taken every 10th, 20th and 50th deme. While the confidence region is larger for smaller numbers of samples, we get a very accurate result even when we have only 9 samples.
}
\label{fig.tdoaBasic}
\end{figure}

We first illustrate this approach on simulated data, where we sample a regular grid (Figure \ref{fig.tdoaBasic}. We simulated a range expansion in a 101 x 101 stepping stone model. In all simulations, we chose the coordinate system such that each deme corresponds to one unit of distance. The start of the expansion is in deme (25,35), indicated by the grey dotted lines in Figure \ref{fig.tdoaBasic}. The direction of the arrows plotted in Figure \ref{fig.tdoaBasic} indicate the sign of the pairwise $\psi$-value, between adjacent samples on a grid, and the thickness of each arrow corresponds to the magnitude of $\psi$. A missing arrow denotes a non-significant $\psi$ value. In Panel \ref{fig.tdoaBasic}A we performed a simulation under an equilibrium isolation-by-distance model. We see that in this scenario, only 11 out of the 60 pairwise comparisons are significant; all of them point towards the corners and are due to the boundary effects of the simulations.  The red ellipse is a 95\% confidence ellipse of the 
inferred origin. Under the isolation-by-distance model, this is located in the center of the population, illustrating that the TDOA approach will yield an answer even if there is no expansion has occured, so it is important to first test if an expansion has actually occured. From Panels B-D we see that  the expansion signal is clearly portrayed by the directionality indeces and we get high confidence in the estimated origin. In fact, the confidence is so high that the ellipse is barely visible in Panel B, but confidence decreases when we reduce the number of samples. Furthermore, the see from Panels C and D that the origin is slightly biased towards the center of the population. This is again due to a boundary effect, and goes away if we take all samples at least 10 demes away from the boundary of the population.

\begin{figure}[!ht]
\begin{center}
\includegraphics[width=6in]{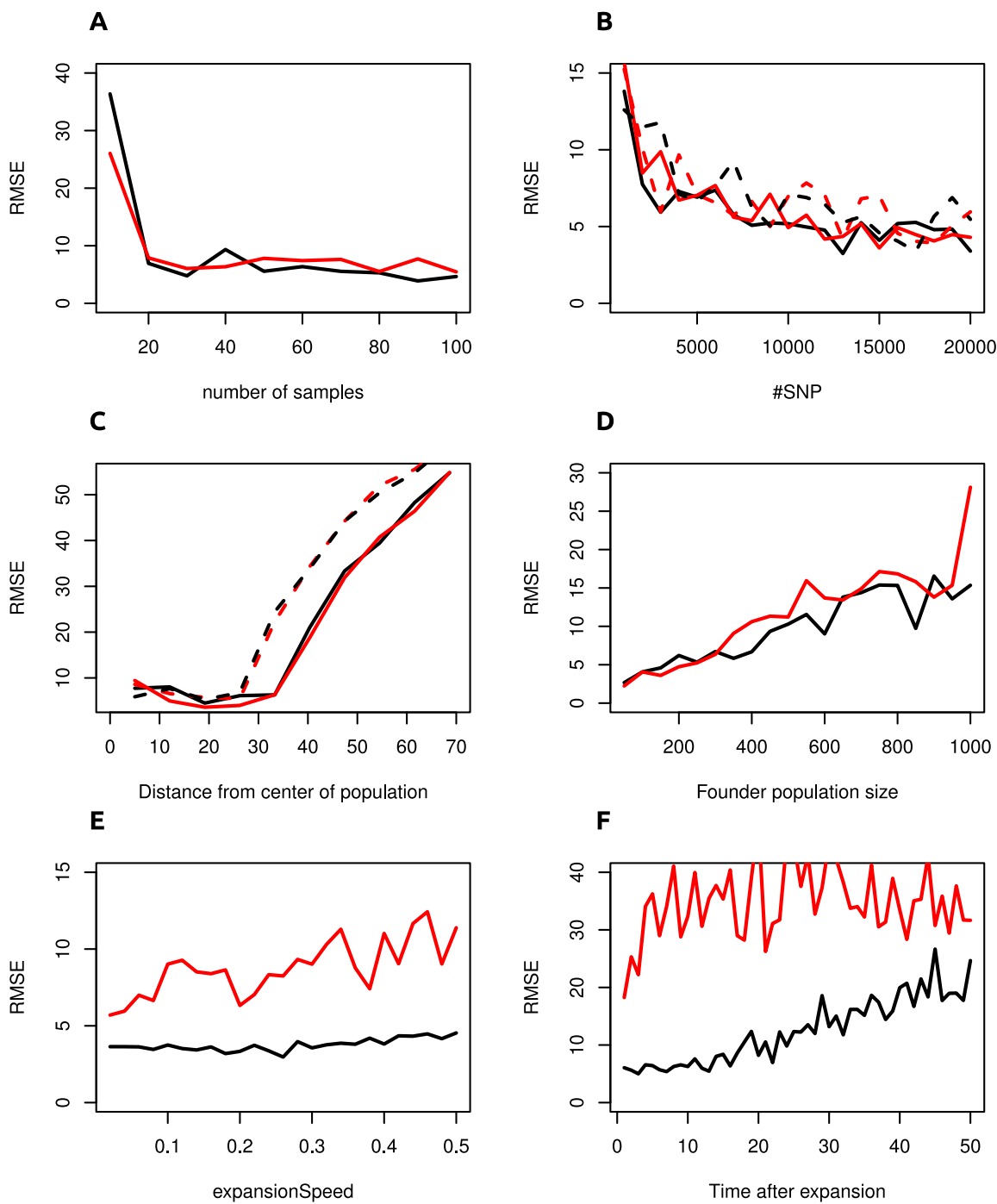}
\end{center}
\caption{
{\bf Performance of TDOA method.} We present the root mean squared errors of our TDOA method (black) compare it with the method of Ramachandran et al. 2005 (red). Samples taken on a grid ware represented by full lines, whereas dashed lines denote samples that were taken from random coordinates in the simulated region. Our method is superior when the expansion occured slowly or when it finished some time in the past; but the method perform very similar for recent, fast expansions.
}
\label{fig.powertdoa}
\end{figure}

To assess the properties of this method more systematically, we report root mean squared error (RMSE) under several scenarios (Figure \ref{fig.powertdoa}).  We also compare our method to the method of Ramachandran et al. \cite{ramachandran2005}, who used linear regression of the heterozygosity on the distance to candidate origins. Their inferred origin of the expansion is the point with the highest associated regression coefficient, conditional on the slope of the regression curve being negative. Most data in Figure \ref{fig.powertdoa} was simulated with a fairly rapid expansion; the time between subsequent expansion events was set to 0.001 coalescence units, so that the complete expansion was completed in 0.13 coalescence units. This speed is roughly equal to the out-of-Africa expansion of humans. For these parameters (Figure \ref{fig.powertdoa}A-D) the two methods have similar performance, with only marginal improvements in how the methods perform with different amounts of data. We find that with adequate 
numbers of samples and data, the RMSE for both method is around four, with less than one distance unit of difference between the two methods. Overall, the ideal amount of data for this method lies around 20 diploid samples and 7,000 independent SNP. Increasing the amount of data beyond this level will not substantially improve performance. For the set of simulations with increasing numbers of SNP, we also tested the effects of sampling on a grid versus taking samples from random locations. It might be assumed that the latter scenario is closer to a realistic sampling scheme. Interestingly, we found only negligible differences, indicating that the sampling locations are only a minor issue unless the sampling locations are very skewed (e.g. a transsect sample).

Changing the position of the origin has little effect on the RMSE for the first 30 distance units, indicating that we have good accuracy if the origin is not close to the border. If there is an expansion that started outside the sampled range, the method will perform significantly worse. This has two causes: first, we would expect it to be easier to infer the origin if it lies in the middle of the sample, as compared to an origin that is far away from all samples. This part also explains the difference between samples taken on a grid and random samples: In the grid, the corners are systematically sampled (since we force a grid sample to be there), whereas in many random samples there may be fewer samples on one side of the origin than on the other, resulting in a loss of accuracy. A second factor resulting in reduced accuracy are again boundary effects, which skew the effect of the expansion if it happened close to the boundary.

We focus our attention on the effect of the parameters of the expansion (\ref{fig.powertdoa}C-F): The number of founders (Figure \ref{fig.powertdoa}d) has an almost linear effect on the estimation accuracy. Fewer founders imply a stronger founder effect and hence a stronger signal of expansion \cite{slatkin2012}, which makes the origin easier to detect. We find the biggest difference in how our method performs in comparison to the Ramachandran method when slowing down the expansion, or when we want to detect an expansion that occured some time in the past. Interestingly, our method performs at almost the same accuracy independent of the expansion speed, whereas the accuracy of the Ramachandran method declines faster. Also, we find that the heterozygosities approach equilibrium soon after the expansion has finished (\ref{fig.powertdoa}F), whereas the shared alleles used by $\psi$ keep the signature of the range expansion for much longer.

\subsection*{Adding Environmental Complexity}
The previous section assumes an idealized population in a homogeneous habitat. In practice, however, habitats are heterogeneous and barriers to gene flow and pathways of expansion are often very important. In the following sections, we show how our method performs in slightly more complex scenarios. First, we allow demes with different population sizes. While we kept the mean size of demes the same, we followed \cite{Wegmann2006} in drawing deme sizes from a gamma distribution. A second important feature not present in the previous simulations are barriers to dispersal that affect both the initial expansion and gene flow following the expansion. Depending on the species, these barriers may correspond to rivers, mountains or even roads. We illustrate how we can use graph algorithms and the directionality index to identify them. Finally, we explore the case with an expansion starting from multiple points, and how we can infer the coordinates of the origins.

\subsubsection*{Heterogeneous Population Sizes}
The effect of variance in deme size on demographic expansions was explored extensively in a simulation study by Wegmann et al. \cite{Wegmann2006}. They found that heterogeneous populations have a higher rate of population differentiation between demes, and predicted that detecting range expansion would be more difficult because of the increased noise. Our simulations confirmed this prediction but only if there is substantial variation in deme size. We found that heterogeneity in deme size has little effect if the variance in deme size is low, with RMSE only differing slightly from the case with equal deme sizes. A variance of 0.5 in deme size, for example, corresponds to a size difference of around two orders of magnitude between the largest and smallest deme. But the average RMSE for the location estimate only increased to 5.43, compared to 4.57 in a comparable scenario without variation in deme size. However, this value corresponds to some kind of ``tipping point'': when we further increasing the variance 
in deme size,
 some deme sizes will become effectively zero in size and this greatly reduces the accuracy of the location estimate, indicating that significant barriers to gene flow make our location estimate less precise.

\subsubsection*{Barriers}
\begin{figure}[!ht]
\begin{center}
\includegraphics[width=6in]{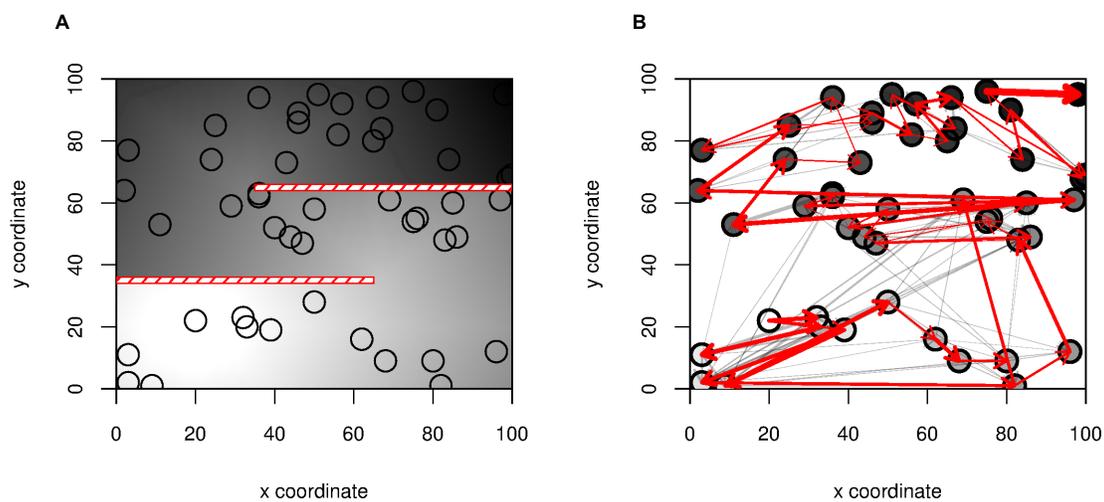}
\end{center}
\caption{
{\bf Identifying complex patterns of migration.} We simulated data on a S-shaped habitat with two impermeable barriers (Panel A) The darkness of the shading is proportional to the arrival time of the expansion, which began in deme (20,20). Black circles correspond to locations sampled. In Panel B we show the inferred pairwise directionality, with all edges remaining after thinning the graph shown in grey, and a maximum spanning tree in red. We also show the inferred ordering of the samples as a color gradient of the samples from light (closest to origin) to dark. The barriers can be identified from panel B by the absence of any indication of gene flow across the barriers and by examining the ordering of the samples.
}
\label{fig.barriers}
\end{figure}

We can use pairwise directionality indices to gain information about colonization paths, i.e. the corridors through which the population expanded. We approach this problem by interpreting a set of pairwise directionality indices as a directed graph, where the samples correspond to the vertices and the directionality indices correspond to the edges connecting the samples. To achieve a visual representation of the graph, we apply graph algorithms to remove some of the edges. In particular, we use the transitive reduction algorithm \cite{aho1972}. A transitive reduction finds the graph with the fewest edges that retains the connectivity of the original graph. That is,  if there is an edge from vertex $v_1$ to vertex $v_2$, but there is also an indirect path from $v_1$ to $v_2$ via another vertex $v_3$ (i.e. $psi$ is significantly negative between $v_1$ and $v_3$ as well as between $v_3$ and $v_2$), then the edge from $v_1$ to $v_2$ is removed from the graph. A further reduction can be obtained by computing a 
maximum spanning tree\cite{korte2008}, which reduces the graph to $n-1$ edges, where $n$ is the number of samples. The maximum spanning tree representation should be able to identify major migration paths, and does not cross strong barriers of gene flow (Figure \ref{fig.barriers}). Furthermore, we can obtain an ordering of all samples by simply summing all $\psi$ values that sample is involved in:
\begin{equation}
 \psi_i = \sum_{j\in\text{samples}}\psi_{ij}\text{.}
\end{equation}
The lowest value of $\psi_i$ among the samples denotes the sample that was taken closest to the origin, and the highest value of $\psi_i$ is the sample furthest along the expansion. In Figure \ref{fig.barriers}B we show that both the maximum spanning tree and the ordering are useful tools and able to identify the barriers.

\subsubsection*{Multiple Origins}
\begin{figure}[!ht]
\begin{center}
\includegraphics[width=6in]{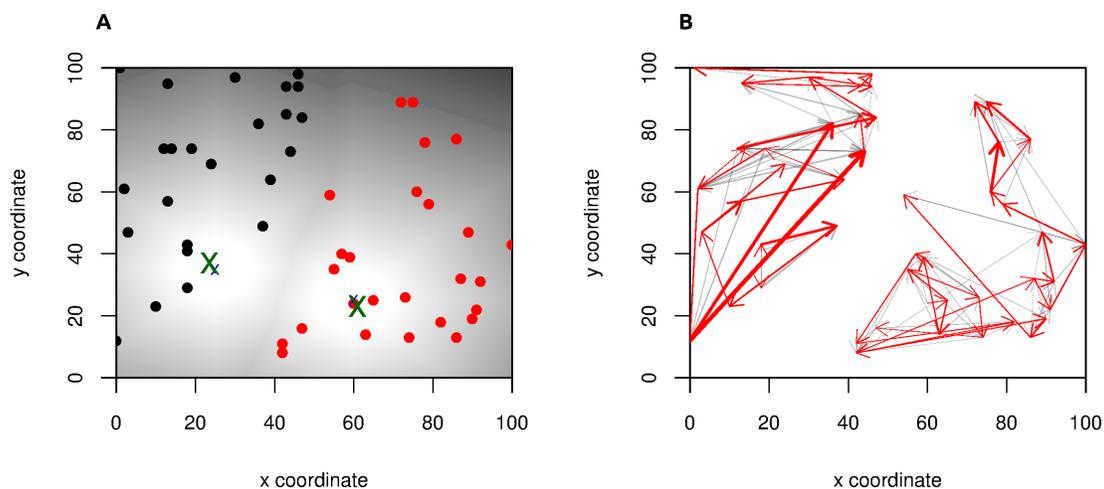}
\end{center}
\caption{
{\bf Detecting multiple origins.} Panel a: We simulated two expansions that originated at the same time from origins indicated by the blue crosses. The color gradient in the background corresponds to the time of colonization time of each deme. We address the problem of inferring the origin of multiple expansions using a two-step procedure. First, we cluster the samples into discrete clusters (red and black, respectively) and then estimate the expansion signal and origins independently for the clusters, resulting in high accuracy for both origins (green X). The right panel shows the inferred migration patterns after a transitive reduction (grey/red arrows) and a maximum spanning tree (red arrows).
}
\label{fig.multiOrigin}
\end{figure}

Range expansions may have more than one source. A classical example is the colonization of Central Europe after the last glacial maximum. Species with Southern European refugia in the Balkan Penisula, Italy or the Iberian peninsula followed the receding glaciers and explain many biogeographical pattern we observe today \cite{schmitt2007}. Our method extends in a straightforward manner to such expansions. The key idea is to first find samples that were predominantly colonized from a single origin, and then estimate the position of that origin independently. There are several ways to assign sampled individuals to clusters corresponding to a single origin. In classical studies, often mtDNA haplotypes were used for this purpose (e.g. \cite{hewitt1999,taberlet1998}), but programs such as STRUCTURE \cite{pritchard2000} or simple clustering based on the observed polymorphism frequencies may yield more accurate results. In our simulations, a simple K-means clustering algorithm was able to correctly identify the 
number of clusters in all cases, even when the two founder populations were drawn from the same original population. The resulting estimates of the origins are slightly less precise than with a single origin (Figure \ref{fig.multiOrigin}), but that is to be expected as there are fewer samples contributing to the location estimate for each origin. Also, the estimates were worse when the two origins were close together.

\subsection*{Application}

\subsubsection*{Human Diversity}
We applied our method to a data set from 55 human populations from the Human Genome Diversity Panel and HapMap III \cite{fumagalli2011,cann2002,theinternationalhapmap3consortium.2010}. We calculated $\psi$ and its standard error for all pairs of populations and transformed this into a Z-score. As expected from a data set with several hundred thousand loci, the vast majority of comparisons were highly significant, with a median absolute Z-score of 28.1, and a mean absolute Z-score of 41.9 across all comparisons made. Globally, we could make out four major clusters of populations: i) Africans, ii) Europeans and Pakistani, iii) East Asians and iv) Native Americans. Overall, every one of the 450 comparisons made between a population in Africa and Non-African showed evidence for gene flow out of Africa, confirming the out-of-Africa hypothesis. Within Africa, we found all comparisons to be 
significant, and all pairwise $\psi$ values agreeing on a single origin of the expansion. The San people were the only population that had positive $\psi$ values when compared to all other populations, indicating that they are closest to the origin of humans. They are followed by the Biaka- and Mbuti-pygmies, which are have negative $\psi$ values when compared to the San. This is followed by the southern Bantu sample, and a cluster consisting of Yerubans, Luhya, Mandenka and Northern Bantu, each having a negative $\psi$ for other previously mentioned populations, and positive scores for all other populations. The African populations furthest away from the origin were the Maasai and Mosabite, the latter being very distinct from the sub-Saharan populations.

The closest populations outside Africa are the Bedouin and Palestinian populations, both from the Middle East. The third Middle Eastern population present in our data, however, the Druze people, fall in a large cluster containing almost all European, Pakistani and Indian populations. Within Europe, the three Italian population samples all have non-significant $\psi$ scores, but are found to be more ancestral to the other European populations. They are followed by the French and French-Basque, which also cannot be distinguished, and the Orcadian, Adygei and Russians. In Pakistan, we find the Makrani to be the most ancestral population, followed by the Brahui and Balochi, Sindhi, Kalash and Burusho. It is noteworthy that this list corresponds to their distances from Africa, with the exceptions that the Brahui and Balochi are switched, and the Hazara are not in the main Pakistani cluster, but rather form a distinct group with the Uygur. Besides the Uygur, all other East Asian populations form a single 
large cluster with very little resolution. Clearly distinct from this cluster are the Papuans and Melanesians, which are very similar with asymmetry between these two populations($\psi=0.0019$, $SE\psi=9.2e-4$, $Z=-2.05$). They are closer to the Africans than to the East Asians, but further away than the Pakistani and European populations. 

Finally, Native American populations form a distinct cluster, which are strongly separated from all other populations. Within the Native American populations, we find evidence of a North to South colonization pattern with Pima being closest to the Eurasian populations, followed by the Maya and Colombians. The most distant populations are the South American Karitiana and Surui, which have a nonsignificant pairwise $\psi$ between them.

We also tested our ability to infer the origin of humans using the TDOA approach. As continents most likely act as strong migration barriers, we did not use the TDOA approach on the entire HGDP data set. Instead, we applied our method to the data set of Henn et al \cite{henn2011} which contains 30 African populations. We estimate an origin of the Human expansion at 30$^\circ$ S 13$^\circ$ E, which lies in central South Africa, closest to the San sample locations at 28.5$^\circ$ S 21$^\circ$ E and 22$^\circ$ S 20$^\circ$ E, respectively. 

\section*{Discussion}
In this paper, we introduced a new statistic, the directionality index $\psi$. We then showed that $\psi$ can be used to reject an equilibrium isolation-by-distance model, and we used it both to characterize  a range expansion and estimate its origin. Although we have focused on range expansions, $\psi$ is sensitive to other deviations from symmetric migration. While a range expansion might be a plausible explanation in many cases, alternative scenarios such as a source-sink population structure or a large differences in effective population sizes should also be considered. One of the main advantages of the simplicity of the directionality index is that the assumptions - and limitations of the approach are easy to discern: the directionality index is zero if the 2D-SFS is roughly symmetric about the diagonal. This is certainly true under most models considered in theoretical studies, such as island and stepping stone models, particularly as the boundary conditions in the latter are typically chosen such that 
the 
model is symmetric. The directionality index can be used to determine how appropriate these models are for a given data set. If $\psi$ differs from zero then care should be taken in applying methods that are based on these theoretical models. On the other hand, if $\psi$ is close to zero, we can interpret this as justification for using the powerful theoretical results for these models \cite{durrett2008}.

In this regard, the directionality index can be seen as a ``first step'' analysis that can be computed very easily, is able to answer very broad questions about our data and then be used as a guide to which more complicated parametric models might be employed, e.g. in an Approximate Bayesian Computation\cite{beaumont2002,wegmann2010} or dadi \cite{gutenkunst2009} framwork. We have also shown how we can introduce the physical location of the samples in our inference framework. In many cases, natural populations are better described by a continuous distribution \cite{rosenberg2005,guillot2009}, and as we show in the TDOA analysis, using a simple statistic together with the physical locations can result in a quite powerful method. Our approach is also different from most other methods dealing with spatial data in that it explicitly assumes a non-stationary population. In this paper, we link the ancestral demographic process of a range expansion to the observed patterns of genetic diversity. While the effect of 
the expansion on $F_{ST}$ appears to be quite small, our $\psi$ statistic can be used to distinguish between equilibrium and non-equilibrium models. Finally, we also show how we can extend our method to deal with complications. Whereas the TDOA analysis is not robust to large barriers of gene flow, interpreting the pairwise $\psi$ statistics as a graph can unmask important details of a species' history.

\subsection*{Simulation Results}
We find that the directionality index $\psi$ is well suited to distinguishing between isolation-by-distance and range expansion when demes are sufficiently far apart and the range expansion is recent and occurs at a fast rate. These restrictions are not surprising. Geographically close demes will be genetically more similar, regardless of their history, and historical processes should therefore be harder to distinguish. That a recent expansion is easier to detect than an older one is also easily explained by the eventual convergence to equilibrium isolation-by-distance pattern, and similarly, a rapid range expansion leaves less time for genetic drift to blur the patterns created by the range expansion. Lastly, increasing the amount of data will increase the power to distinguish asymmetric from symmetric processes as each single SNP contributes only a little information about the history of dispersal. In all cases, our $\psi$ statistic outperforms $\Delta H$.  From the analyses under the stepping-stone 
model we see one of the main differences between $\psi$ and genetic distance measures, such as $F_{ST}$. In an isolation-by-distance model, $F_{ST}$ will increase with distance, but $\psi$ will not deviate from zero as the distance between the sampled locations increases. Again, this makes sense intuitively: The number of shared genetic variants decreases with distance, and hence $F_{ST}$ increases. However, this reduction in shared polymorphisms is symmetric, and hence will have no effect on $\psi$. The pattern is different in the model of a population expansion: when comparing with a sample from the origin of expansion, both $F_{ST}$ and $\psi$ increase with distance. The signal diminishes, when migration rates are high, however. This is apparent from Panel D in Figure \ref{FigDistance}, where $\psi$ is zero for the first ten demes. Here, migration had enough time to undo the effect of the range expansion in the demes that are further away from the origin.

In the origin estimation section, we find that we can get surprising accuracy with relatively little data. 20 samples with around 10,000 SNP yield accurate estimates. This amount of data indicates that our method is not applicable to mtDNA or microsatellite data, but it should be applicable to transcriptome data, which can be assembled for many non-model organisms. It is also worth noting that the error does not go to zero even with larger amounts of data. This can have several reasons. A big contribution is likely from the linearity assumption we made for the TDOA approach. $\psi$ does not increase perfectly linearly with distance, and especially the boundaries of the simulated region may introduce a considerable bias. A second, more subtle point is the algorithm we use; whereas least-squares is very easy to use and yields good results, other optimization algorithms might reduce the RMSE. A third explanation is that genetic processes are simply very noisy, and we require much more data to obtain better 
results. Our results also show that using heterozygosity to infer the origin of an expansion is largely similar to our statistic for recent, fast expansions. 

We demonstrated how our method can be adapted to incorporate more complex models. We showed that small differences in deme sizes have little influence on our ability to estimate the origin. If however, the habitat is very heterogeneous our method becomes less accurate. This implies that some care must be taken when for example analysing endangered species with very patchy habitats, or species whose dispersal distance is much lower than the size of a local population, or when analyzing global patterns of diversity If there are strong barriers to the gene flow, the TDOA method should not been applied, as its central assumption that $\psi$ is proportional to physical distance is violated. In that case, while it is not possible to infer an origin that is distinct from the samples, it is nevertheless possible to find the sample that is closest to the origin, which in many cases might suffice to support or reject a hypothesis. Also, we have shown that we can apply graph algorithms to get a representation of the 
migration pattern that leads to meaningful interpretation.

\subsection*{Human Genetic Diversity}
When analysing the human data sets, we found that i) $\psi$ scores are correlated with distance and ii) if population $i$ is closer to Africa than population $j$, then $\psi(i,j)$ is in most cases negative, a pattern that is expected under a model of expansion from Africa. As explained previously, the directionality index depends not only on the two population compared but also on the history of the other populations. We find the South African San people to be the population closest to the origin of humans both using the TDOA method and when interpreting all pairwise directionality indeces. This supports the interpretation that the origin of modern humans is somewhere in Southern Africa \cite{tishkoff2009, henn2011}. Another interesting result is that the Melanesian and Papuan samples, while very similar, show positive $\psi$ values 
when compared to other East Asian populations, but the directionality index is negative when compared to the Pakistani, European and African populations. This is consistent with a ``two-wave'' model of colonization of South-East Asia, with a first wave consisting of present-day Papuans and Melanesians, and a second wave consisting of the present day Chinese populations\cite{rasmussen2011}.


\section*{Methods}
\subsection*{Simulations}
We implemented a simulator that performs continuous time coalescent simulations on a discrete stepping stone model \cite{malecot1950, kimura1964} of finite size. We assumed that the backward migration rates were equal between all pairs of adjacent demes and that the boundaries were reflecting. We used a modified version of the expansion model of \cite{slatkin2012}, where an expansion is modeled with a one-generation bottleneck of reduced size. In our backward-in-time framework, this corresponds to moving all lineages present in a deme being colonized to a randomly chosen neighboring deme. We introduce a founder effect by adding additional coalescence events according to the appropriate backward Wright-Fisher transition probability (Page 62 in \cite{wakeley2009}). Unless noted otherwise, all expansions were done with a founder size of 200. Once the final deme is reached, an regular island model coalescent is run where each island corresponds to a founder population (in most simulation, the number of 
islands is one).

Throughout this paper, we simulated unlinked SNPs using an importance sampling scheme. After generating 1,000 gene trees, we calculate the appropriate multi-dimensional site frequency spectrum, where each sampled population corresponds to a dimension. We can then draw SNPs with replacement from this site frequency spectrum.

The parameters used for the majority of power simulations are as follows: We simulated on a 101 x 101 stepping stone model, with deme coordinates starting at (-50,50) at the lower left corner and (50,50) in the upper right corner. Each deme exchanges migrants to the neighboring demes to the north, south, east and west at rate $M=1$. For the power simulation, we sampled a single diploid individual each from two colonies at (-25,-25) and (-25,25). For the TDOA simulations we simulated one individual each from a deme on a quadratic grid between (-30,-30) and (30,30), with 36 samples in total. This corresponds to a distance of 12 demes between any two sampled demes. We usually generated 1,000 independent coalescent trees and then used importance sampling to generate 100,000 SNP from the population, conditioning on them being shared between at least two of the samples. In the case of a range expansion, the standard point of origin was set to (-15,-25) and the expansion occurred at a rate of one expansion event 
every 0.001 coalescence units, with the expansion being observed 0.13 coalescent units after it started, where coalescent units are measured on the time scale of a local deme. These parameters were chosen to roughly correspond to the human out-of-Africa expansion.
The directionality index $\psi$ and $F_{ST}$ were calculated in Python; for $\psi$ we used equation (\ref{dIndexAlt}), and $F_{ST}$ was estimated using Reynold's estimator \cite{reynolds1983}. Note that these are only baseline parameters, and exploring the effect of changing these parameters is the main purpose of most of our power simulations.

To generate data for the 1D stepping stone model analyzed in Figure 1, we simulated a 201 x 1 habitat, with scaled migration rates $M=1,10$ between adjacent demes. Sampling was done in demes $-i/2$ and $i/2$, with the center deme having coordinate 0. In case of range expansions, the expansion started in deme $-i/2$.

SNP ascertainment may influence our results, because most ascertainment schemes favor high frequency alleles in the populations where the ascertainment was performed. To assess the effect of ascertainent bias on the value of $\psi$, we performed simulations in an isolation-by-distance model with samples at coordinates (0,0), (10,0), (20,0), (30,0), (40,0), (50,0) as well as (0,10) and (15,10) and then computed $\psi$ between the (10,0) and (20,0) sample. We then simulated ascertainment by selecting a set of population, and rejection sampling SNP so their 1D-SFS followed a Beta(2,4/3) distribution, which roughly matches the SFS in the HGDP data set and is very different from the expectation without ascertainment bias. If $\psi$ differs significantly from zero, then we know that ascertainment is important. Results are given in Figure S1; ascertainment is important if it is performed in one of the populations that we calculate $\psi$ for. However, the effect of ascertainment is negligible if the population we 
calculate $\psi$ for are different from the ascertainment population, even if the ascertainment population is much more closely related to one population compared to the other.
\subsection*{Estimating the origin of a range expansion}
We use a time-difference of arrival (TDOA) approach \cite{Gustafsson2003} to estimate the origin of a range expansion. TDOA was originally used in naval navigation during the Second World War, and is currently widely used to solve localization and navigation problems. It is based on the assumption that a single source emits a signal that decays with increasing distance from the origin. For range expansions, this signal can be thought of as the mean allele frequency of shared alleles. At the origin, the allele frequency is expected to be lowest \cite{slatkin2012} and to increase approximately linearly with distance. However, since we do not know the allele frequency at the origin, we have to use the indirect approach by comparing pairs of populations. To be precise, if we know that shared alleles have a lower frequency at point $S_i$ compared to point $S_j$, then we know that $S_i$ is closer to the origin than $S_j$. If the habitat is two-dimensional, however, this does not tell us the direction of the 
expansion. Let 
$||S_i,S_j ||$ denote the Euclidean distance between two points $S_i$ and $S_j$. Then, 
\begin{equation}
 ||S_i,O|| - ||S_j,O|| \approx v \psi_{i,j},
\end{equation}
where $O$ denotes the unknown origin $\psi_{i,j}$ is the directionality index between samples $S_i$ and $S_j$ and $v$ is a constant that links space to allele frequency (i.e how much does the allele frequency change per unit of space). In words, $\psi_{i,j}$ is approximately proportional to the difference of the distances $||S_i,O||$ and $||S_j,O||$ (see also Figure \ref{fig.tdoaBasic}). We assume that the sampling locations of $S_i$ and $S_j$ are known without error, and that $\psi_{i,j}$ can be estimated from genetic data, along with its sample variance $\operatorname{Var}(\psi_{i,j})$. We estimate the variance by doing 1,000 bootstrap replicates on the SNP. The unknowns that remain are the coordinates of the origin $O$ and the proportionality constant $v$. To infer these parameters, we solve for $\psi$, subtract $\psi$ from the equation and sum over all pairs of samples:
\begin{equation}
 \left (\hat{O},\hat{v} \right ) = \underset{O,v}{\operatorname{argmax}}  \sum_{i<j} \frac{1}{ \operatorname{Var}(\psi_{i,j})}\left ( \frac{||S_i,O||-||S_j,O||}{v} - \psi_{i,j} \right ).
\end{equation}
In most biological application, space will be two-dimensional and therefore we can make this equation more explicit by writing $O=(x,y)$ and $S_i=(x_i,y_i)$. Then,
\begin{equation}
 \left (\hat{x},\hat{y},\hat{v} \right ) = \underset{x,y,v}{\operatorname{argmax}}  \sum_{i<j} \frac{1}{ \operatorname{Var}(\psi_{i,j})}\left ( \frac{1}{v} \left ( \sqrt{(x_i-x)^2+(y_i-y)^2} -  \sqrt{(x_j-x)^2+(y_j-y)^2} \right ) - \psi_{i,j} \right ).
\end{equation}
The variance terms correspond to weighting terms; terms where $\psi$ has a high variance are weighted down, whereas terms where we can infer $\psi$ with high accuracy are given a larger weight. We can then find a solution to this equation using nonlinear least squares.


\section*{Acknowledgments}
This research was supported in part by NIH grant R01-GM40282 to MS. We would like to thank Laurent Excoffier, Kelley Harris and Josh Schraiber for helpful discussions.


\section*{Figure Legends}


\begin{figure}[!ht]
\begin{center}
\includegraphics[width=6in]{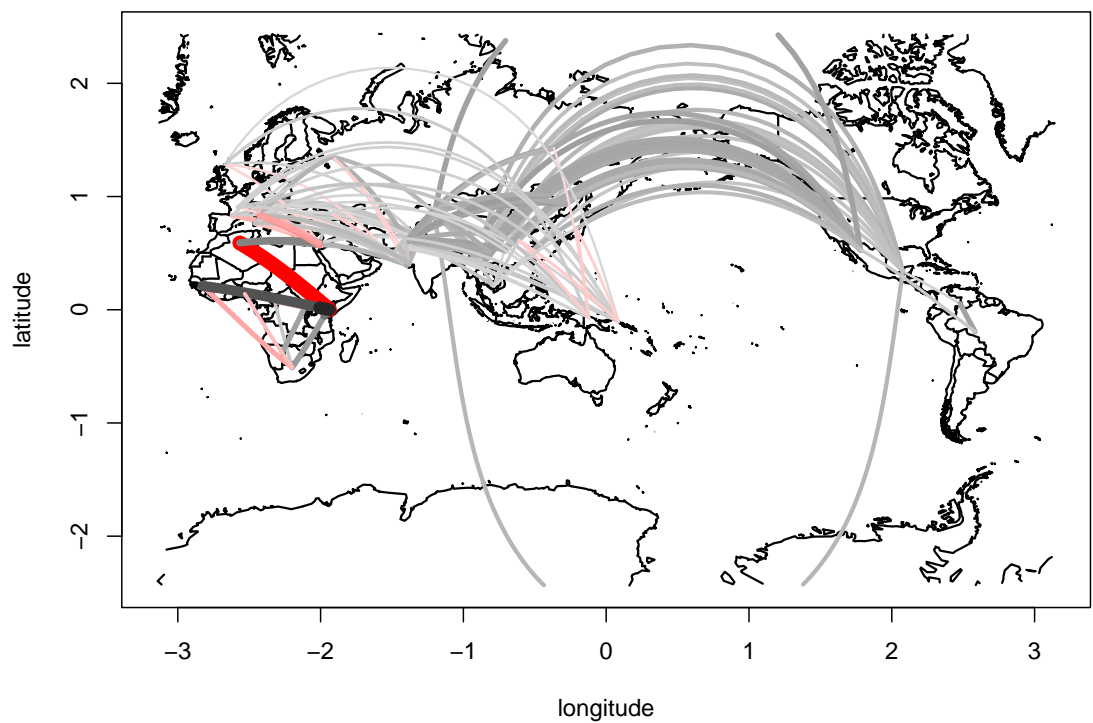}
\end{center}
\caption{
{\bf Inference of human migration routes.} The figure shows a visual representation of the pairwise directionality indices between human populations in HGDP and HapMap. Each line corresponds to the pairwise $\psi$ statistic, with thicker and brighter lines corresponding to higher values. Grey and red lines denote eastward and westward migration, respectively. Lines with an absolute Z-score below 5 were omitted.
}
\label{fig.humans}
\end{figure}


\begin{figure}[!ht]
\begin{center}
\includegraphics[width=6in]{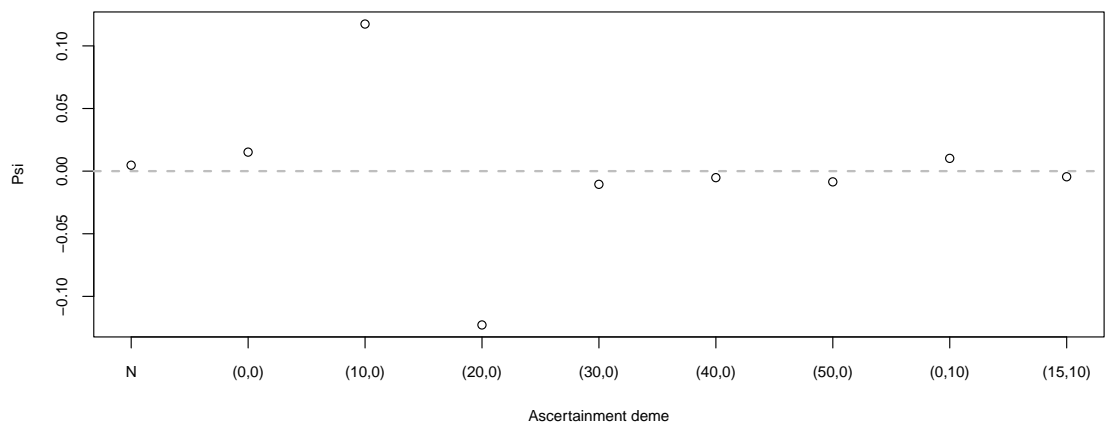}
\end{center}
\caption{
{\bf Effect of ascertainment bias on $\psi$.} We show the effect of strong ascertainment bias in different demes given on the x-axis on $\psi$ calculated between samples taken from coordinates (0,10) and (0,20). N = no ascertainment. Ascertainment has very little effect if it is performed in a deme that is not used in the comparison.
}
\label{fig.ascertainment}
\end{figure}

\section*{Tables}

\end{document}